\documentstyle [12pt]{article}

\title{
{\vspace{-1.5cm} \normalsize
\hfill \parbox{38mm}{NHCU-HEP-94-22 \\
                     MS-TPI-94-11 }   }\\[25mm]
Study of the Decoupling
 of Heavy Fermions in a $Z_2$ Scalar-Fermion Model}

\author{ Lee~Lin \\
Department of Physics,\\
National Chung Hsing University,\\
250 Kuo Kuang Road,\\
  Taichung 40227, Taiwan, ROC}
\newcommand{\be}{\begin{equation}}
\newcommand{\ee}{\end{equation}}
\begin{document}
\maketitle

\begin{abstract} \normalsize
According to one-loop perturbation theory, fermions whose
masses are totally generated from Yukawa couplings
do not decouple in the heavy mass limit. We investigate this issue
nonperturbatively in a 4-dimensional $Z_2$
scalar-fermion model with staggered fermions.
Our data at intermediate
and stronger Yukawa couplings on $8^4$ and $12^4$ lattices suggest the
nondecoupling of heavy fermions as predicted from one-loop calculation.
However, at the strongest Yukawa coupling where a possible multi-critical
point may come into play, we cannot be conclusive.
\end{abstract}

\section{ Introduction }

The decoupling theorem \cite{ACC}
says that when a particle has a mass much higher than the physical
scale, it will have very small influence (e.g.
radiative corrections) on the ``physical world". However,
it can be easily seen at one-loop level that the contribution
of the particle to renormalized quantities of other particles will
not be suppressed by its huge mass if its mass is generated from the
Yukawa coupling, i.e. if its mass is generated through the mechanism
of spontaneous symmetry breaking (SSB). This is the
so-called phenomenon of nondecoupling.

Nondecoupling of heavy fermions in theories with SSB
has been discussed in several papers \cite{LC,DEC,AP}.
Until last year, all arguments were within (one-loop) perturbation theory
(except in \cite{AP} where large-N expansion was used).
Nondecoupling beyond one-loop was still not clear and should be explored.

Recently, we studied this issue in a nonperturbative way
in a $U(1)$ scalar-fermion model with explicit mirror-fermions \cite{LL1}.
There, we ignore the gauge field assuming that this approximation
will not change the picture qualitatively.
Hence, the scalar field is our ``physical world".
The relevant quantity to measure to decide decoupling or not is the
ratio of the two renormalized Yukawa couplings: One is the
usual renormalized coupling defined as the renormalized (mirror-)fermion mass
divided by the renormalized vacuum expectation value (VEV); the other is
obtained from the fermion-fermion-scalar 3-point vertex function.
In the weak coupling regime, one-loop calculation shows that the ratio
of the two is very close to one. As the bare Yukawa coupling gets
stronger, the renormalized Yukawa coupling defined from the mass-to-VEV ratio
is no longer a good definition for the renormalized coupling. There, we
need to define the renormalized coupling from the appropriate 3-point
vertex function. If the ratio of the two Yukawa couplings (denoted
by $R$) still remains 1.0, we take it as the indication for the
nondecoupling of the heavy (mirror-)fermion.

\vspace{0.2cm}
Our main results in \cite{LL1} are:

\vspace{0.2cm}
\noindent
(i) Up to the strong Yukawa coupling regime,
 the ratio $R$ for the mirror-fermion is always equal to 1.0
within the error, indicating that
the heavy mirror-fermion does have its renormalized
Yukawa coupling proportional to the mass and
will not decouple from the transverse component of the scalar field.
Thus one-loop picture survives the strong Yukawa coupling
limit. The idea of decoupling the mirror partners
by giving them large masses does not work.

\vspace{0.2cm}
\noindent
(ii) Since the action of the model has a symmetry between
fermion and mirror-fermion ,
 it is obvious that the heavy fermion itself will not decouple either.

\vspace{0.2cm}
\noindent
(iii) Our data also show that all doublers decouple as expected.

\vspace{0.2cm}
\noindent
(iv) We think that these conclusions also apply to
the $SU(2)$ version of the mirror-fermion model
because its qualitative behaviour appears to be similar to
that of the $U(1)$ model \cite{FLMMPTW}.

\vspace{0.2cm}
\noindent
(v) We conjecture that in other scalar-fermion
models, heavy fermions do not decouple either.

\vspace{0.2cm}
In order to confirm our conjecture mentioned in (v),
we carried out Monte Carlo simulations on a 4-dimensional $Z_2$
scalar field theory coupled to the staggered fermion.
We now report on our results in this letter.

\section{Action and Renormalized Quantities}

The lattice action of the 4-dimensional $Z_2$ scalar-fermion model
with staggered fermions is
$$
S=-2\kappa\sum_x\,\sum_{\mu=1}^4\,\phi_x\,\phi_{x+\mu}+\lambda\sum_x(\phi_x^2
-1)^2+\sum_x\phi_x^2
$$
\be \label{eq01}
+{1\over 2}\sum_x\,\sum_{\mu=1}^4\,\bar\chi_x\,\eta_{x,\mu}\,(\chi_{x+\mu}
-\chi_{x_\mu})+ S_{\rm int}
\ee
where $\phi_x$ is the scalar field with only one real component,
$\chi_x$ and $\bar\chi_x$ are the one-component staggered fermion fields,
$\eta_{x,\mu}$ being the phase factor associated with staggered fermions,
$\eta_{x,1}=1$ and $\eta_{x,\mu}=(-1)^{x_1+\cdots+x_{\mu-1}}$ for
$2\le\mu\le 4$, $S_{\rm int}$ represents the term for the Yukawa interaction.
In this report, we study the model with the so-called
``non-overlapping" Yukawa coupling introduced in \cite{BHHK}. Hence,
$$
S_{\rm int} = G\sum_x\bar\chi_x\Phi_x\chi_x\,\, ,
\,\, \Phi_x\equiv{1\over 16}\sum_{x'\in hypercube(x)}\,\phi_{x'}
$$
where the set of sites $x'$ is obtained by
$$
{x'}_\mu = 2 \Bigl[{x_\mu\over 2}\Bigr] + \delta_\mu\,\, ,\,\,
\delta_\mu = 0, 1\,\, ,
$$
and $[.]$ denotes the largest integer part of the arguement.
The boundary condition is chosen such that it is periodic for the scalar field
and the spatial directions of the fermion fields, and antiperiodic
along the temporal (4th) direction of the fermion fields.
Since each species of staggered
fermions will generate four Dirac fermions in the continuum with opposite
chiralities, the physical spectrum will be vector-like.

The above action has the following discrete symmetry:
\be \label{eq02}
\phi_x\rightarrow -\phi_x\, ,\, \chi_x\rightarrow i(-1)^{\sum_\mu x_\mu}\chi_x
\, ,\,\bar\chi_x\rightarrow i(-1)^{\sum_\mu x_\mu}\bar\chi_x\,\, .
\ee

For the study of the decoupling of heavy fermions, the relevant phase (called
the FM phase from now on) is the
one where $Z_2$ symmetry in eq.(\ref{eq02})
is spontaneously broken. Notice that we do not have
the bare fermion mass term in the action. This means that the renormalized
fermion mass will be totally generated by the Yukawa coupling in the FM phase.
Definitions of various renormalized quantities there are as follows.

The renormalized scalar mass $m_R$ and the scalar
wavefunction renormalization constant $Z_\phi$ are defined as
\be \label{eq03}
{Z_\phi\over \hat p_4^2+m_R^2}\equiv \lim_{p_4\to p_{4\rm min}}\langle
\tilde\phi(p_4)\tilde\phi^*(p_4)\rangle_c\,\,
\ee
where subscript $c$ means the connected part of the propagator,
$$
\tilde\phi(p_4)={1\over \sqrt{\Omega}}\sum_x\,e^{-ip_4x_4}\phi_x\,\,
,\,\,\hat p_4^2=4\, {\rm sin}^2({p_4\over 2})\,\, ,
$$
$\Omega$ is the total number of lattice sites, $x$ means $({\vec x},x_4)$,
and $\tilde\phi^*(p_4)$ is the complex conjugation of $\tilde\phi(p_4)$.
In our simulation on a finite $L^4$ lattice, we set $p_{4\rm min} = 2\pi/L$.
The renormalized VEV is defined as
\be \label{eq04}
v_R\,\equiv\,{\langle|\phi|\rangle\over\sqrt{Z_\phi}}\,\,
,\,\, |\phi|\,\equiv\,{1\over\Omega}\sum_x\,\phi_x\,\, .
\ee

Similarly, the renormalized fermion mass $\mu_R$ and the fermion
wavefunction renormalization constant $Z_F$ are defined as
\be \label{eq05}
{Z_F\over \mu_R+i\,{\rm sin}(p_4)}\,\equiv\,\sum_y\, e^{-ip_4y_4}
\langle\chi_y\bar\chi_0\rangle\,\equiv\,Z_F(\tilde\Gamma_R(p_4))^{-1}\,\,
\ee
where the spatial 3-momentum is set to zero, $p_4$ is chosen to be
$\pi/L$ for the fermion fields, and $\tilde\Gamma_R(p_4)$ is the
renormalized fermion 2-point vertex function in momentum space.
 The renormalized Yukawa coupling $G_R$ is defined as
\be \label{eq06}
G_R\,\equiv\,{\mu_R\over v_R}\,\, .
\ee

The relevant quantity to tell
decoupling from nondecoupling of heavy fermions in this $Z_2$ scalar-fermion
model of staggered fermions with nonoverlapping Yukawa coupling is still
the ratio of the 3-point renormalized Yukawa coupling and the one defined
in eq.(\ref{eq06}). The 3-point
renormalized Yukawa coupling coupled to the scalar field is defined as
\be \label{eq07}
G^{(3)}_R\,\, \delta_{k,-p+q}
= - {{\hat k}^2_4+m_R^2\over Z_F\,\sqrt{Z_\phi}}\,\tilde{\Gamma}_R(p_4)\,
\,G^{(c)}\, \tilde{\Gamma}_R(q_4)\, \ ,
\ee
where $k_4$, $p_4$, $q_4$ are
the $4$th components of the momenta of scalar field, fermion and
anti-fermion, respectively, and $\hat k_4^2$ is $4\,\rm{sin}^2(k_4/2)$.
We have set the spatial components of all momenta to zero.
The appearance of the Kronecker-delta above is due to
energy-momentum conservation.
$G^{(c)}$ is the connected part of the $\phi$-$\chi$-$\bar\chi$ 3-point
Green's function and is
\be \label{eq08}
G^{(c)} = \frac{1}{L^4}\,
\sum_{x,y,z}\,e^{-ik_4x_4}\,e^{-ip_4y_4}\,
e^{iq_4z_4}\Bigl<\phi_x\chi_y\bar\chi_z\Bigr >_c.
\ee
In our simulations on $L^4$ lattices we choose
$$
k_4 = {2\pi\over L}\,\, ,\,\,
p_4 = -{\pi\over L}\,\, ,\,\, q_4={\pi\over L}
\,\, .
$$
The ratio $R$ is defined to be $R=G^{(3)}_R/G_R$.

Notice that SSB only occurs in the zero-momentum
mode, the above connected 3-point
Green's function $G^{(c)}$ is equal to the disconnected one, because it is
defined at $k=(\vec 0,k_4)$ with $k_4\ne 0$.
In the $Z_2$ scalar-fermion models, there are no massless
Goldstone bosons in the FM phase since no continuous symmetry is
spontaneously broken.
We simply treat the scalar field itself as the ``physical world" and
assume that the issue of the decoupling of heavy fermions does not
depend on whether the broken symmetry is discrete or continuous.
Notice that reflection positivity for scalar-fermion models with
staggered fermions cannot be proven. We assume that no ghost
particles are present in the spectrum.

\section{ Phase Structure and Numerical Simulation }

The phase structure of the model defined in eq.(\ref{eq01})
at $\lambda=0.01$ has been explored
and was presented in figure 2 of \cite{BHHK}.
At $G=0$, the model reduces to a one-component pure scalar $\lambda
\phi^4$ theory which has FM, symmetric (PM)
 phases and an anti-ferromagnetic (AFM) phase.
At $\lambda=0.01$, the transition point between FM and PM phases occurs
around $\kappa=0.120$ and is a Gaussian fixed point. Due to the symmetry:
$\phi_x\rightarrow(-1)^x\phi_x, \kappa\rightarrow -1\kappa$,
the transition between PM and AFM phases occurs around $\kappa=-0.120$.
As $G$ is gradually turned on, the FM-PM phase transition
occurs at smaller and smaller (and eventually negative) $\kappa$ values
because dynamical fermions favour the FM phase.
This feature has been observed in all scalar-fermion models studied
so far \cite{BHHK,ALL}. As the value of $G$ is further increased,
some new phase, like the ferrimagnetic (FI) phase,
shows up and the phase structure will be (slightly) dependent on the
model \cite{ALL}. For the $Z_2$ scalar-fermion model investigated in
this letter, the FI phase and a new PM phase were found at strong and
infinite $G$ \cite{BHHK}. Around $G=1.7$, there is a point (called
point B in \cite{BHHK}, and will be called as such from now on)
around which four phases (FM, PM, AFM and FI) may coexist.
 We have explored the phase structure
around that point B on $4^3\cdot 8$ lattice, and find that point B
is located around $G=1.7$ and $\kappa=-0.125$, consistent with previous
estimate \cite{BHHK,JB}.

The physically interesting phase transition is the one between the FM and
PM (called S1 in \cite{BHHK}) phases (curve AB in figure 2 of \cite{BHHK}).
It was found to be consistent with a second-order phase transition on
which the cutoff can be removed. We therefore carry out
Monte Carlo simulations at $\lambda=0.01$ in the FM phase, but in the
vicinity of this FM-PM phase transition line.

In the simulations, besides $\lambda=0.01$,
we set $G$ to be 0.3, 0.6, 1.0, 1.2 and 1.8.
The lattices we use are $8^4$ and $12^4$.
We then tune the scalar mass parameter $\kappa$ to have $m_R$ around
1.0 on $8^4$ lattice to reduce finite size effects and
at the same time maintain a reasonable cutoff.

In this $Z_2$ scalar-fermion model, no continuous symmetry is
spontaneously broken in the FM phase, so there is no massless particle
in the spectrum. Finite size effects should be small except on small
lattices where vacuum tunnelings happen. It is well known that on a finite
lattice, the ground state is nondegenerate. This means that there is no
SSB on a finite lattice. The unique ground has a vanishing VEV and is a
symmetric linear combination of two states, which are
peaked at the positive and negative minima respectively.
It is these two states that will converge to the
two degenerate ground states in the thermodynamic limit \cite{GM}.
Once tunnelings happen on a finite lattice, we will be in the
nondegenerate ground state of the system. Finite size effects will be
dominated by an instanton-like equation \cite{GM} and may not be small on
small lattices. If the system is not too close to criticality such that it
is trapped in one of the minima, then finite volume effects will be dominated
by perturbative effects rather than tunneling. In the presence of
dynamical fermions, the precision of our data will
not be good enough. To analyze finite size effects dominated by
tunnelings is therefore too demanding for the moment.
So, when we study the properties of the FM phase of the model
on a finite lattice and eventually extrapolate to get the infinite
volume limit, we actually do not wish to see tunneling events.

Without loss of generality in our simulations,
we always set up the initial condition such that the system starts in the
positive minimum. We define the time-slice average of the scalar field as
$$
\phi_s(t)\equiv\, {1\over L^3}\sum_{\vec x}\phi_{\vec x, t}\,\, .
$$
By observing values of those $\phi_s(t)$, we will know whether
tunnelings happen or not.

The Monte Carlo simulations are performed by the unbiased
Hybrid Monte Carlo method \cite{HBMC}.
Therefore, the fermions have to be doubled by taking the
adjoint of the fermion matrix for the second species.
(The fermionic part in eq.(\ref{eq01}) is given for a single species
of staggered fermions.) We will have eight degenerate Dirac fermions
in the continuum limit. The number of leapfrog steps per molecular dynamics
trajectory was chosen randomly between 3 and 10. The step size was
tuned to maintain an acceptance rate around $75\%$. The necessary
inversions of the fermion matrix were done by the conjugate
gradient iteration, until the residuum was smaller than
some small value times the length square of the input vector.
We find that this value has to be $10^{-12}$ on the $8^4$ and $12^4$
lattices. We use Creutz observable $exp(-\delta H)$ to decide whether
the system has equilibrated or not \cite{MC}
where $\delta H$ is the difference between the new and old Hamiltonians
in the Hybrid Monte Carlo update. In equilibrium, we should have
$$
\langle e^{-\delta H}\rangle = 1\, \, .
$$

\section{ Conclusion and Discussion }

Our numerical data are presented in table 1.
Our data on the renormalized fermion mass and bare VEV at $G=$ 1.2 and
1.8 agree with those published in \cite{BHHK}. We also find that the
renormalized fermion mass has the mean field behaviour $\mu_R=G\langle
|\phi|\rangle$. The renormalized fermion mass presented here
can be as high as $700$ to $800$ $GeV$ at strong Yukawa couplings.
(The physical scale is set by $v_R = 246$ $GeV$.)
At all points where we did simulations, no vacuum tunneling events were
observed. We think this is partly due to the fact that dynamical fermions
tend to increase the height of the barrier between the two minima. It is
also because Hybrid Monte Carlo is a local updating algorithm. (In the
presence of dynamical fermions, we cannot use cluster algorithms to
perform global updates.) Thus
it is not easy for the system to tunnel even on an $8^3$ spatial lattice.

According to table 1, data on $R$ from $G=0.1$ to $1.2$
are all consistent with 1.0 within errors, indicating a universal
behaviour of the nondecoupling of heavy fermions as expected.

However, at $G=1.8$, $\kappa=-0.04$, data on $R$ on $8^4$ lattice
(i.e. point g in table 1) is clearly smaller than 1.0
by several standard deviations, while data on
$12^4$ lattice are too noisy to tell.
Therefore, we cannot be certain whether heavy fermions are still coupled
to the scalar field at $G=1.8$. At the moment, we would like to say that
there could be a possibility that heavy fermions do decouple
as the decoupling theorem says, although their masses are
totally generated from Yukawa coupling. (In the action of our model,
we do not have a bare fermion mass term.)
This possibility exists because points f, F, g, G in table 1 are
lying slightly to the right of point B and may very well be affected by
it. Since point B might be a multi-critical point and may have,
in principle, renormalization
properties different from those dictated by the
Gaussian fixed point at $G=0$, it should not be too surprising that
heavy fermions decouple in the vicinity of point B.
Besides, if we stay at $G=1.8$ and keep reducing the value of $\kappa$, we
will be approaching the phase transition line between the FM and FI phases.
This FM-FI transition line is the place where bare VEV is still nonzero.
Thus, it should not be a physically relevant phase transition line.

 At this point, we would like to conclude that at least
along the phase transition line between the FM and PM phases
from $G=0.1$ to $G=1.2$ where the system is still governed
by the infrared stable Gaussian fixed point at $G=0$,
heavy fermions whose masses are totally generated
by the Yukawa coupling do not decouple. One-loop picture is
qualitatively correct throughout this region. Although data presented in
this letter are at $\lambda=0.01$, we believe that the above conclusion
holds at all values of $\lambda$ between zero and infinity.
As we go to even stronger Yukawa coupling where a possible multi-critical
point may come into play, the decoupling of heavy fermions remains a
possibility. However, our present data cannot give a conclusive signal
for decoupling. Whether heavy fermions really decouple or not there
depends on the properties of that possible multi-critical point.
Due to limited computer resources, this issue
will be left for the future study.

\section{ Acknowledgement }

I would like to thank T. W. Chiu, G. M\"unster and Z.-Y. Zhu for discussions.
I also thank G. M\"unster for his critical reading of this manuscript.


\newpage
%
\begin{table}[tb]
\caption{  \label{t1} Our Monte Carlo data are presented here.
Points with lower-case letters are obtained on $8^4$
lattice while points E and F are on $12^4$ lattice.
Each point has about 30000 molecular dynamics
trajectories for equilibration, and around 100000 to 300000
for measurements. But point G has only 30000 trajectories for
measurements.
}
\begin{center}
\begin{tabular}
{|c|r@{.}l|r@{.}l|r@{.}l|r@{.}l|r@{.}l|r@{.}l|r@{.}l|r@{.}l|}
\hline
&\multicolumn{2}{c|}{$G$} & \multicolumn{2}{c|}{$\kappa$}
&\multicolumn{2}{c|}{$v_R$}&\multicolumn{2}{c|}{$m_R$}
&\multicolumn{2}{c|}{$\mu_R$}
&\multicolumn{2}{c|}{$G_R$} & \multicolumn{2}{c|}{$G^{(3)}_R$}
&\multicolumn{2}{c|}{$R$}
\\
\hline
 a & 0&1 &  0&129 & 0&600(3) & 0&578(8) & 0&115(1)
   & 0&191(2)&  0&205(43) & 1&07(9) \\
\hline
 b & 0&3 &  0&128 & 0&942(3)  & 0&928(11) & 0&525(2)
   & 0&557(2)&  0&546(60)  & 0&98(7)  \\
\hline
 c & 0&6 &  0&110 & 0&654(2)  & 0&90(1)   & 0&709(4)
   & 1&083(4)&  1&10(10) & 1&02(8)  \\
\hline
 d & 1&0 &  0&080 & 0&496(2) & 1&27(3) & 0&876(2)
   & 1&766(6)&  1&88(17) & 1&06(9)  \\
\hline
 e & 1&2 &  0&060   & 0&437(2) & 1&45(3)  & 0&907(2)
   & 2&08(2)&  2&07(21)  &  0&99(9) \\
\hline
 E & 1&2 &  0&060   & 0&430(2) & 1&34(14)  & 0&906(3)
   & 2&11(2)&  2&17(66)  &  1&03(20) \\
\hline
 f & 1&8 &  0&00    & 0&362(2) & 2&29(6)  & 1&142(2)
   & 3&15(4)  &  2&97(44) &  0&94(5) \\
\hline
 F & 1&8 &  0&00 & 0&368(4)    & 2&21(54) & 1&140(2)
   & 3&10(8)  &  2&2(1.2) &  0&71(40) \\
\hline
 g & 1&8 &  -0&04    & 0&296(1) & 1&37(3)  & 0&766(2)
   & 2&59(2)  &  2&15(24) &  0&83(5) \\
\hline
 G & 1&8 &  -0&04    & 0&275(7) & 1&47(36)  & 0&768(7)
   & 2&79(10)  &  2&9(1.4) &  1&03(46) \\
\hline
\end{tabular}
\end{center}
\end{table}
\end{document}